\documentclass[final,5p,times,twocolumn]{elsarticle} 

\usepackage{graphicx}

\usepackage{amssymb}

\journal{Nuclear Instruments and Methods A}

\usepackage{subfigure}

\begin{document}

\begin{frontmatter}



\title{Status and Objectives of the Dedicated Accelerator R\&D Facility "SINBAD" at DESY}


\author[A]{U. Dorda, B. Marchetti, J. Zhu, F. Mayet, W. Kuropka, T. Vinatier, G. Vashchenko, K. Galaydych, P.A. Walker, D. Marx, R. Brinkmann, R. Assmann}
\author[B]{N.H. Matlis, A. Fallahi, F. X. Kaertner} 
\address[A]{DESY, Hamburg, Germany}
\address[B]{CFEL, Hamburg, Germany}
\begin{abstract}
We present a status update on the dedicated R\&D facility SINBAD which is currently under construction at DESY. The facility will host multiple independent experiments on the acceleration of ultra-short electron bunches and novel, high gradient acceleration methods. The first experiment is the ARES-experiment with a normal conducting 100\,MeV S-band linac at its core. We present the objectives of this experiment ranging from the study of compression techniques to sub-fs level to its application as injector for various advanced acceleration schemes e.g. the plans to use ARES as a test-site for DLA experiments in the context of the ACHIP collaboration. The time-line including the planned extension with laser driven plasma-wakefield acceleration is presented. The second initial experiment is AXSIS which aims to accelerate fs-electron bunches to 15\,MeV in a THz driven dielectric structure and subsequently create X-rays by inverse Compton scattering.
\end{abstract}

\end{frontmatter}

\section{Introduction}
The Helmholtz foundation has identified accelerator R\&D as one of its core tasks. In line with this objective, DESY is currently setting up the dedicated, long term accelerator R\&D facility "SINBAD" in the premises of the old DORIS accelerator complex.

\begin{figure}[htb]
	\centering
	\includegraphics[width=\columnwidth]{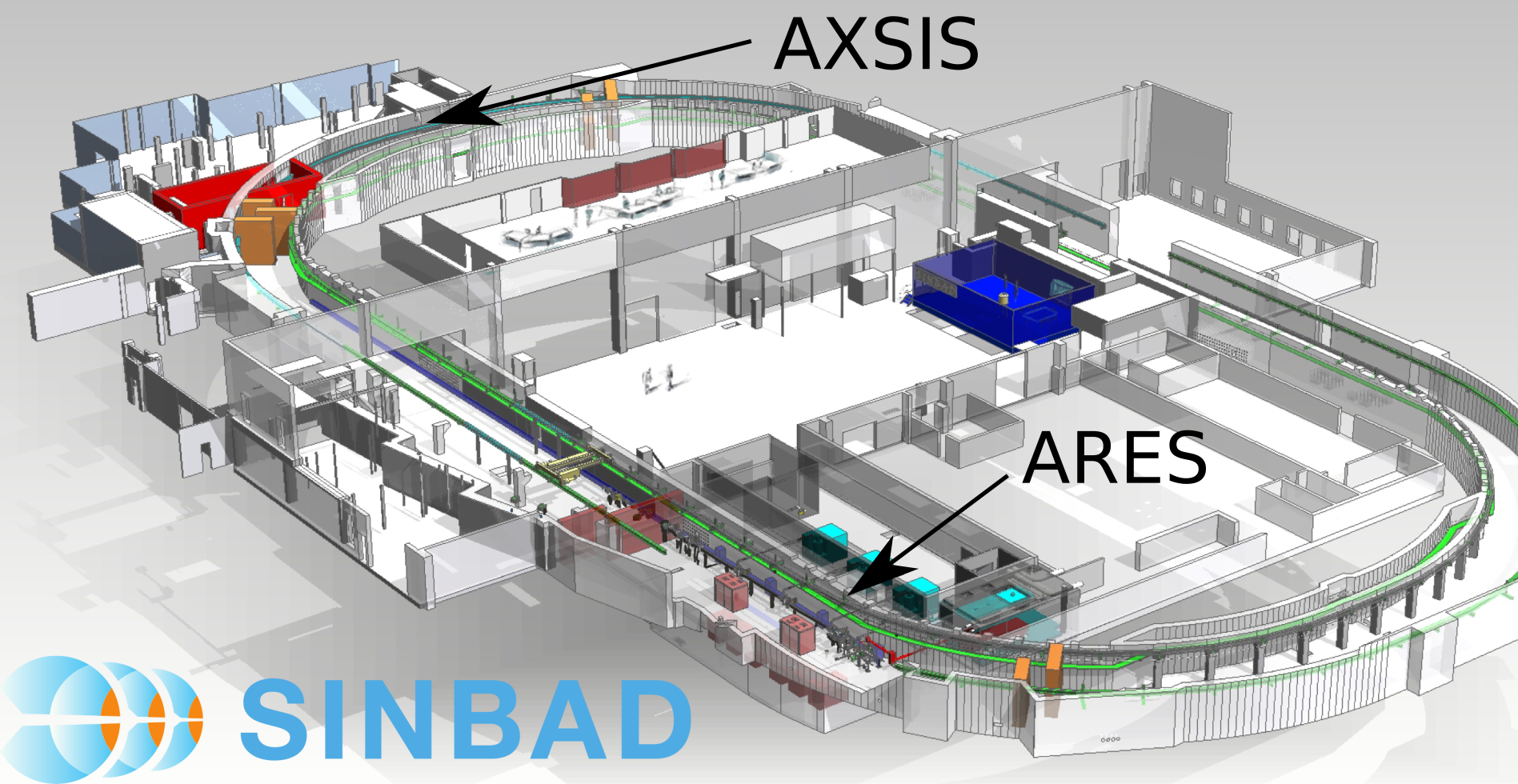}
	\caption{Rendering of the SINBAD facility with the two initial experiments indicated}
	\label{fig:sinbad-3d}
\end{figure}

After 40 years of successful operation, the DORIS accelerator at DESY was shut down at the end of 2012. In the past two years the old accelerator has been removed, the building refurbished, and the installation of the required technical infrastructure is currently finishing. Conveniently located at the heart of DESY, the existing 300\,m long race-track shaped tunnel, features two 70\,m long 5-9\,m wide straight sections and two arcs (Fig. \ref{fig:sinbad-3d}).

The SINBAD project is the framework for all R\&D activities in this area and intends to set up multiple independent experiments in ultra-fast science and high gradient accelerator modules. The first objective includes topics like synchronization R\&D, fs to as-regime electron beams, high brightness beams from RF photo injectors as well as sophisticated beam shaping and will be studied in the first of the two initial experiments, the ARES-linac. Moreover, SINBAD will perform experiments on novel acceleration techniques by applying various different laser-driven approaches with external or internal electron beam injection. This will on the one hand side be done with dielectric structures in the context of the ACHIP collaboration \cite{ACHIP-rasmus} with DLAs at ARES and with THz driven dielectric loaded structures at AXSIS. Finally, laser-driven plasma acceleration as second step behind the ARES linac will be studied in close collaboration with the University of Hamburg and in the context of the Helmholtz wide collaboration on plasma acceleration "ATHENA".

\section{The AXSIS experiment}
One of the two initial experiments, the 'AXSIS' project \cite{AXSIS-eaac}, is a ERC-synergy grant funded collaboration of 4 groups aiming to develop a THz-laser driven, all-optical, compact X-ray source. The millimeter scale of the driving field promises to offer a favorable compromise between conventional accelerators and other advanced accelerators like plasma-based or laser-driven devices with significantly smaller structure dimensions. 

While one part of the team is pushing the amount of achievable THz (300\,GHz) power towards the required levels, various different THz-based guns are designed \cite{arya-guns}. Initial tests of a prototype THz-gun were performed and produced keV level acceleration of electrons using micro Joule, single-cycle THz pulses. While the performance must still be significantly increased with the next guns, the experimental results are now well understood \cite{Vashchenko-THzgun}. With the final gun design, pC level electron bunches will be extracted from a photo-cathode and accelerated to $\approx$ 800\,keV.

In order to not let the space charge effect blow up the bunch, the linac will be located right after the gun. The linac is made from a dielectric loaded waveguide of $\approx$ 10\,cm length \cite{nanni-thzlinac}. The dielectric loading allows to slow down the phase-velocity of the THz-pulse. In the final stage, the electron beam will be focused and collided with an optical laser to create X-rays.

With the corresponding laser labs being constructed right now and the tunnel area infrastructure being installed, we aim to start experiments in the final AXSIS site, in one of the arcs of the SINBAD facility, in fall 2018.

\section{The SINBAD-ARES experiment}
The ARES-linac \cite{marchetti-ares2017} is a S-band (2.998\,GHz) linac which accelerates electron bunches to 100\,MeV while compressing them to fs-length. These ultrashort bunches can then subsequently be used for experiments.
\subsection{The linac setup}
\begin{figure}
	\centering
	\includegraphics[width=\columnwidth]{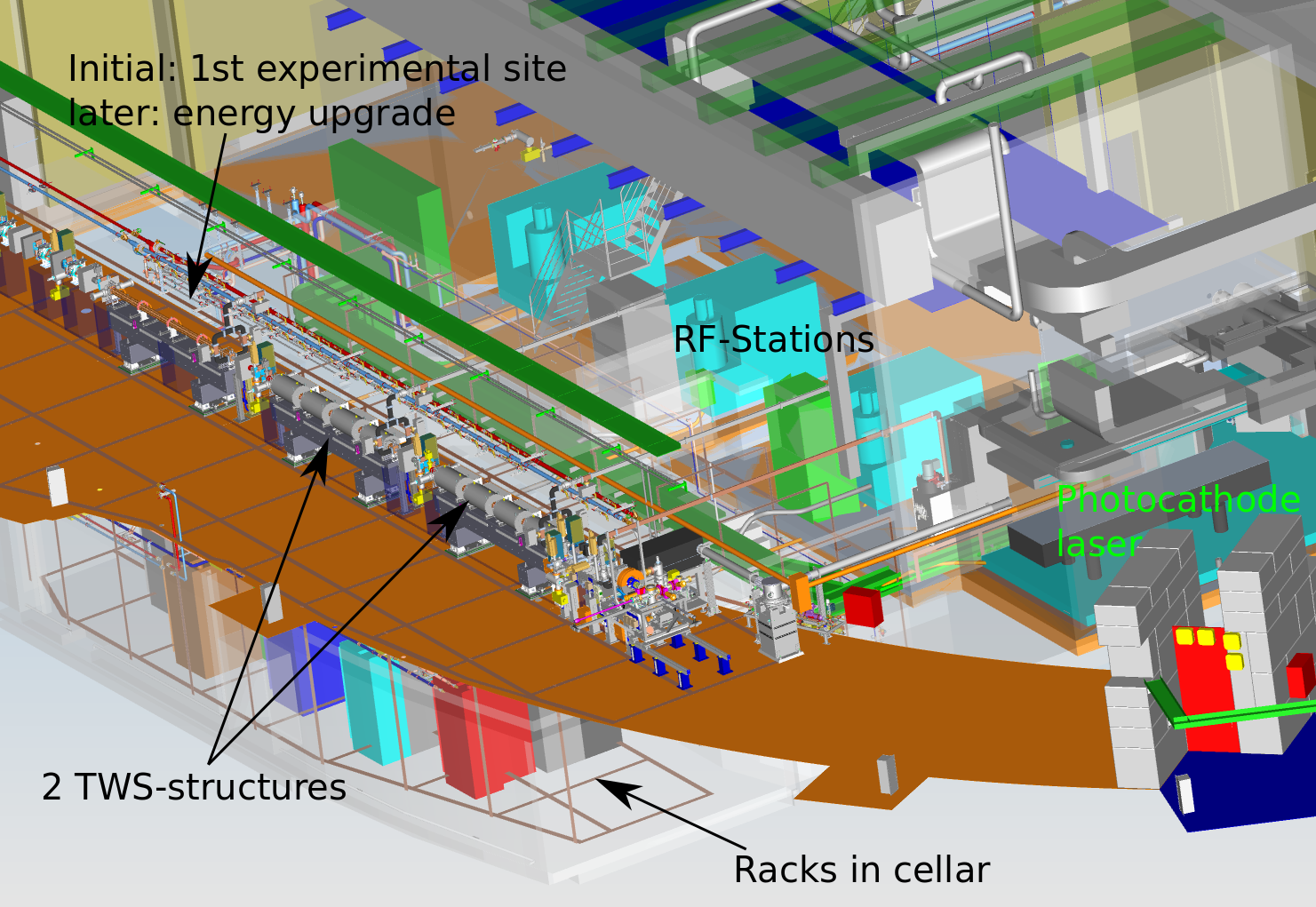}
	\caption{Rendering of the ARES linac showing the beam line in the tunnel as well as the associated infrastructure (RF-stations, photo-cathode laser and rack room)}
	\label{fig:sinbad-ares-cad}
\end{figure}

A CAD rendering of the linac section can be seen in Figure \ref{fig:sinbad-ares-cad}. The electron bunches of typically 0.3-30\,pC are created and accelerated in a 1.6 cell RF-gun (modified version of the one used at DESY-REGAE \cite{regae}) to 5\,MeV. The electrons are extracted by photoemission by a 1\,mJ Yb doped laser at 257\,nm with a variable FWHM pulse length between 180\,fs and 10\,ps. During the first commissioning phase, the first solenoid will be located $\approx 0.4m$ after the photo-cathode. A first iteration of a design for an additional solenoid right at the RF-gun exit has already been developed and is planned to be installed together with a bucking coil in 2019 leading to a reduction of the minimal achievable transverse emittances. Several beam diagnostic devices including screens, Faraday cups etc and a spectrometer dipole, cover about 2.5\,m until the start of the first of two traveling wave linac structures (TWS). These 4.2\,m long TWS with external RF-dump are surrounded by solenoids for emittance compensation. Each one will be powered by a dedicated 45\,MW RF station and could - if operated on-crest - provide 75\,MV energy gain. In practice one or both will be operated off-crest,to compress the bunch to fs-length \cite{marchetti-beammanipulation, Zhu-sim, zhu-compression}. The compression will either be done via velocity bunching in the first TWS or via a magnetic bunch compressor after the linac or a hybrid compression scheme. The two structures are separated by a about 1.2\,m long intertank section containing the necessary beam diagnostic and orbit correctors. The space reserved for the installation of a third TWS will at the start be used for initial experiments. At the end a dedicated beam diagnostic line allow characterization of the beam and optimization of the linac performance. The linac will run in single bunch mode with up to 50Hz repetition rate. The beam line design as well as the whole layout is - within the restrictions of the given building - designed for minimal arrival time jitter.
At ARES, the linac itself will be the object of studies including e.g. comparison of various compression methods, synchronization R\&D and beam shaping techniques. In parallel it will then be used to provide beams for downstream experiments. The ultra short bunches will be ideally suited for injection into advanced schemes and allow probing their phase space acceptance. 

The installation of the gun stage is currently starting up and beam commissioning is aimed for spring 2018. The linac stage will be added in fall 2018, allowing for first experiments in spring 2019.

\subsection{X-band TDS for 6D tomography}
In collaboration with CERN and PSI, an X-band transverse deflecting structure (TDS) with variable polarization is currently being developed \cite{marchetti-xbandcollaboration2017}. In addition to its main objective to measure the longitudinal beam profile with fs resolution (to study and optimize the compression methods), this novel device will allow to select the streaking direction. Using the data from multiple shots with different streaking angles, tomographic methods will be applied to characterize the complete 6D phase-space. \cite{Marx-xbandtomography}

\subsection{Dielectric structure experiments}
\begin{figure}
	\centering
	\includegraphics[width=\columnwidth]{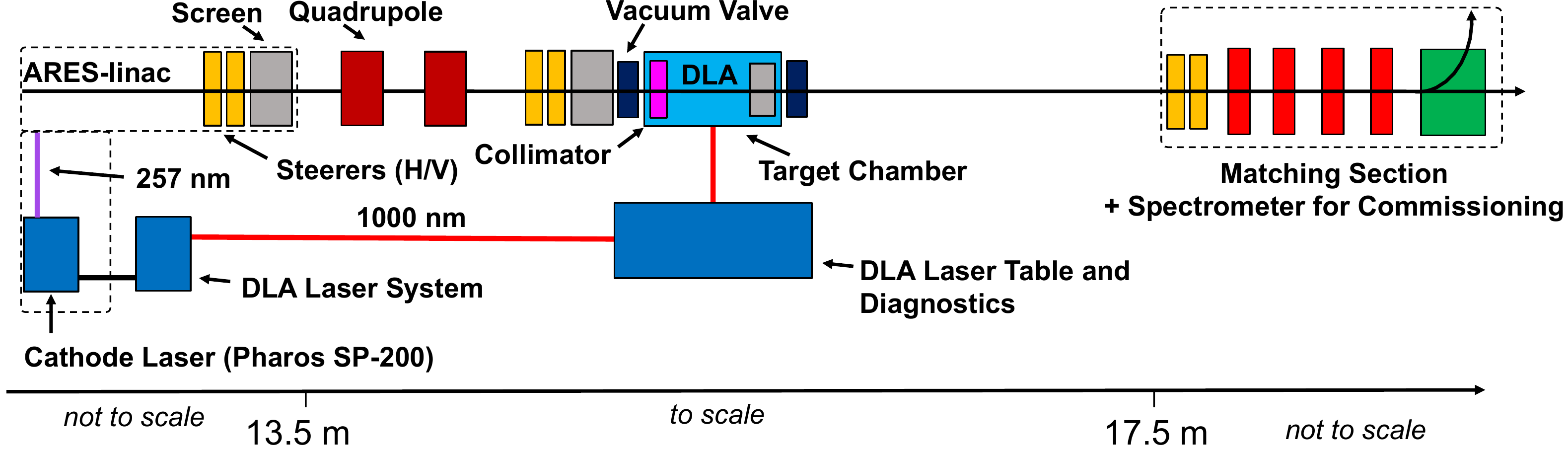}
	\caption{Sketch of the ACHIP experimental area at ARES and the laser split of from the photocathode laser}
	\label{fig:sinbad-achip}
\end{figure}

The SINBAD-ARES linac is one of the intended future test sites for the ACHIP collaboration \cite{ACHIP-rasmus} which studies DLA-type structures. In the first stage it is foreseen to place the experiment at the location which is reserved for a third traveling wave structure (Figure \ref{fig:sinbad-achip}). The required strong focusing of the external beam to the $\mu$-scale structures is provided by the solenoids surrounding the two up-stream TWS structures \cite{Mayet-simulation}. Initially, a part of the photocathode laser will be split off and used to power the DLA structures. It is not yet decided if an optical parametric amplifier will be added to convert the laser from 1\,$um$ to 2\,$um$. Naturally the DLA structure size needs to be adapted accordingly. This setup has the advantage of intrinsic synchronization between the photo-cathode laser- and the DLA drive laser-beams. The relative electron to laser phase jitter at the DLA is hence mainly given by the RF-induced beam arrival time jitter contribution. 

\begin{figure}
	\centering
	\includegraphics[width=\columnwidth]{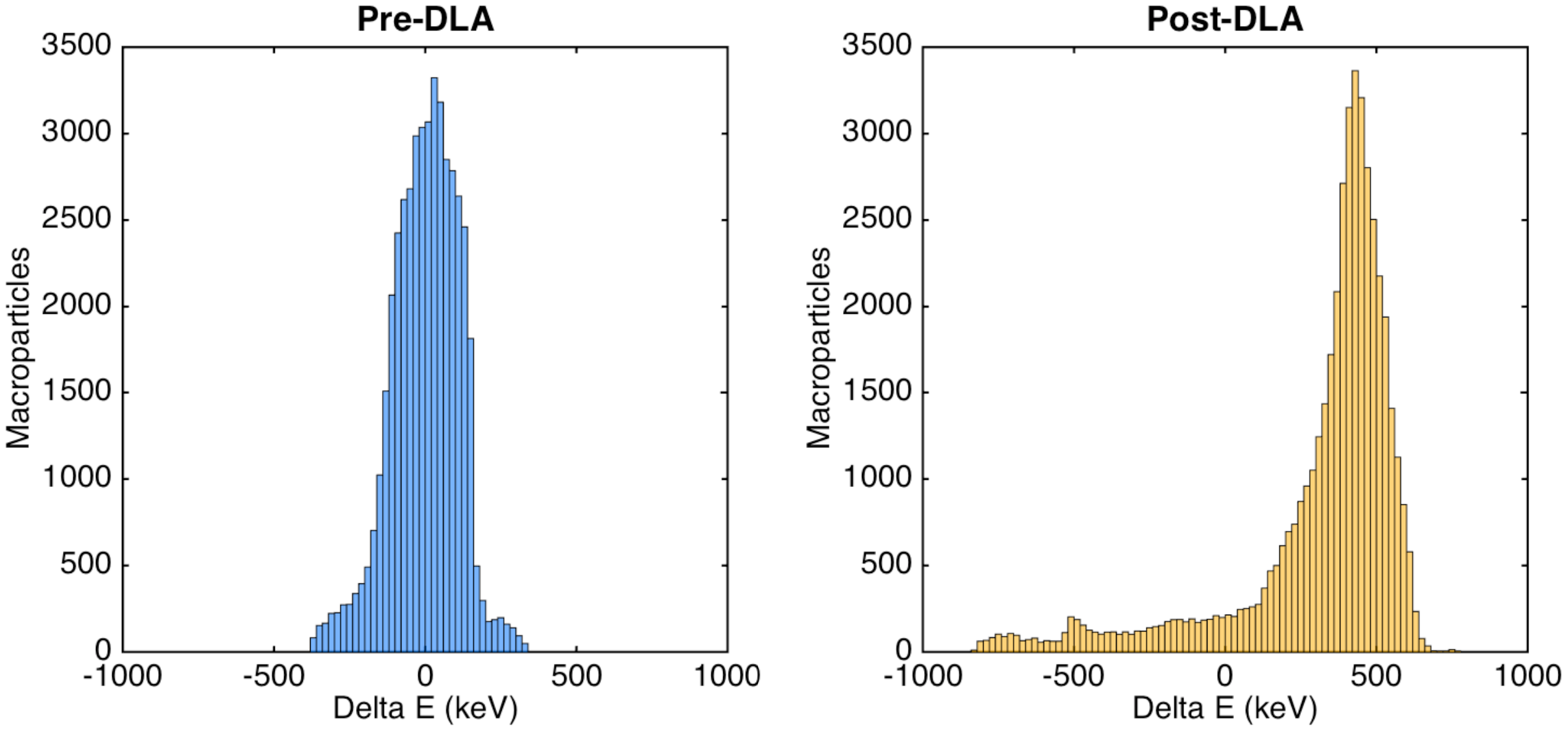}
	\caption{Beam energy distribution before the DLA and after it showing that the shortness of the ARES-linac beam should allow net-acceleration for selected single shots. Depending on the arrival time jitter from shot to shot, the electron to laser phase will vary and thus also cases with net energy loss be observed. Here the ideal case of maximal acceleration is shown.}
	\label{fig:sinbad-achip-energygain}
\end{figure}

In previous acceleration experiments with such structures, the bunch length has been significantly longer than the laser length and thus all electron-to-laser phases have been covered. The shortness of the ARES bunches will allow injecting the bunch only to a limited phase range ($<90$ degree at $2\,um$). While the design of the ARES-linac is optimized to minimize also this RF-induced arrival time jitter, a sub-fs level synchronization will probably still be out of reach. For the shots where the phase happens to be at maximum, net acceleration will be observed (Figure \ref{fig:sinbad-achip-energygain}). In other cases, also net-energy loss or only energy modulation will occur \cite{Kuropka-sim}.

In order to address this issue, in another experiment the DLA-laser beam will be split to not only power the structure but also to micro-bunch the linac beam upstream of the DLA. Similar to the work presented in \cite{searsmicrobunchiung}, this will be done by energy modulating the beam by overlapping it in an undulator and then compressing the bunches in a small magnetic chicane. This creates a number of microbunches spaced at the correct distance for the DLA and intrinsically synchronized to it \cite{Mayet-experiments, Mayet-ipac}.

Furthermore, in another planned experiment the suitability of such structures as transverse deflectors will be studied \cite{Kuropka-simdeflecting}. 

\subsection{THz experiments at ARES}
In parallel to the all optical efforts for THz acceleration in the AXSIS experiments, the linac stage may be tested with the well characterized beam from the ARES linac \cite{Vinatier-sim}. In addition to these acceleration experiments, the high frequency of the THz fields will be used in testing THz streaking experiments or bunch length measurements via the 3 phase method \cite{Vinatier-possibleexp}.

\subsection{Laser plasma wakefield acceleration experiments}
A proposal by 6 Helmholtz centers for a joint laser plasma accelerator research program "ATHENA" has just entered the final approval stage from the Helmholtz foundation. Together with university partners, it aims to push this technology by setting up two flag-ship programs. While studies on laser plasma acceleration of ions will be performed at HZDR in Dresden, the electron activities will be focused at SINBAD. For this purpose, a high power laser lab will be installed in the large central hall in the next few years. The high power laser will then be used to serve multiple experiments. While experiments using the externally injected beam from ARES will be done in one straight section, internal injection will be studied in the second straight section. Like this SINBAD will be a important preparation step towards the implementation of the EUPRAXIA design study \cite{andieupraxia, Svystun-sim}.

\subsection{External access via ARIES TNA}
While SINBAD is a DESY project, access will be possible to external researchers. This will mainly be done in the context of collaborations, but also via the ARIES transnational access program free of charge \cite{aries}. This EU-funded Integrating Activity project aims to develop European particle accelerator infrastructures and contains the task to open research facilities to external users.

\section{Summary/Conclusion}
We have presented the objectives and status of the SINBAD facility and its two initial experiments, AXSIS \& ARES. The installation of the dedicated accelerator R\&D facility is progressing well and experiments on ultra-fast science and advanced acceleration schemes are developed.

\section{Acknowledgements}
\label{sec:acknowledgements}
The authors acknowledge the work of all the DESY technical groups involved in the renovation, planning and construction of the facility. In addition, we would like to mention our collaboration partners at LAOLA, ATHENA \& ACHIP. The AXSIS related research leading to these results has received funding from the European Research Council under the European Union's Seventh Framework Programme (FP/2007-2013) / ERC Grant Agreement n. 609920'. Access via the ARIES-TNA program will partially be sponsored via the European Union’s Horizon 2020 Research and Innovation programme under grant agreement No 730871. ACHIP related activities are partially funded by the Gordon and Betty Moore foundation (GBMF4744).



\bibliographystyle{plain} 
\bibliography{2017-09-25_-_Dorda_-_SINBAD_for_EAAC}

\begin{thebibliography}{10}

\bibitem{aries}
Accelerator research and innovation for european science and society.
\newblock https://aries.web.cern.ch/.

\bibitem{arya-guns}
A.~Fallahi and et. al.
\newblock Short electron bunch generation using single-cycle ultrafast electron
  guns.
\newblock {\em Phys. Rev. Accel. Beams}, 19:081302, Aug 2016.

\bibitem{regae}
Klaus Floettmann and et. al.
\newblock "{REGAE}: {N}ew source for atomically resolved dynamics".
\newblock {\em Research in Optical Sciences}, 2012.

\bibitem{ACHIP-rasmus}
R.~Ischebeck and et. al.
\newblock The accelerator on a chip international program {ACHIP}.
\newblock {\em Nuclear Instruments and Methods in Physics Research Section A:
  Accelerators, Spectrometers, Detectors and Associated Equipment}, 2017.

\bibitem{Kuropka-sim}
W.~Kuropka and et. al.
\newblock Full {PIC} simulation of first {ACHIP} experiment @ {SINBAD}.
\newblock {\em Nuclear Instruments and Methods in Physics Research Section A:
  Accelerators, Spectrometers, Detectors and Associated Equipment}, 2017.

\bibitem{Kuropka-simdeflecting}
W.~Kuropka and et. al.
\newblock Simulation of deflecting structures for dielectric laser driven
  accelerators.
\newblock {\em Nuclear Instruments and Methods in Physics Research Section A:
  Accelerators, Spectrometers, Detectors and Associated Equipment}, 2017.

\bibitem{marchetti-beammanipulation}
B.~Marchetti and et. al.
\newblock Electron-beam manipulation techniques in the {SINBAD} linac for
  external injection in plasma wake-field acceleration.
\newblock {\em Nucl. Instr. and Methods in Physics A}, 829, 2016.

\bibitem{marchetti-ares2017}
B.~Marchetti and et. al.
\newblock Status update of the {SINBAD-ARES} linac under construction at
  {DESY}.
\newblock In {\em Proceedings of the 8th International Particle Accelerator
  Conference}, 2017.

\bibitem{marchetti-xbandcollaboration2017}
B.~Marchetti and et. al.
\newblock X-band {TDS} project.
\newblock In {\em {P}roceedings of the 8th International Particle Accelerator
  Conference}, 2017.

\bibitem{Marx-xbandtomography}
D.~Marx and et. al.
\newblock "new measurement techniques using a novel {X}-band transverse
  deflecting structure with variable polarization".
\newblock {\em Nuclear Instruments and Methods in Physics Research Section A:
  Accelerators, Spectrometers, Detectors and Associated Equipment}, 2017.

\bibitem{AXSIS-eaac}
N.~Matlis and et. al.
\newblock Acceleration of electrons in {THz} driven structures: first steps
  towards {AXSIS}.
\newblock {\em Nuclear Instruments and Methods in Physics Research Section A:
  Accelerators, Spectrometers, Detectors and Associated Equipment}, 2017.

\bibitem{Mayet-ipac}
F.~Mayet and et. al.
\newblock A concept for phase-synchronous acceleration of microbunch trains in
  {DLA} structures at {SINBAD}.
\newblock In {\em Proc. IPAC'17}, 2017.

\bibitem{Mayet-experiments}
F.~Mayet and et. al.
\newblock Simulations and plans for possible {DLA} experiments at {SINBAD}.
\newblock {\em Nuclear Instruments and Methods in Physics Research Section A:
  Accelerators, Spectrometers, Detectors and Associated Equipment}, 2017.

\bibitem{Mayet-simulation}
F.~Mayet and et. al.
\newblock Using short drive laser pulses to achieve net focusing forces in
  tailored dual grating dielectric structures.
\newblock {\em Nuclear Instruments and Methods in Physics Research Section A:
  Accelerators, Spectrometers, Detectors and Associated Equipment}, 2017.

\bibitem{nanni-thzlinac}
E.~Nanni and et. al.
\newblock Terahertz-driven linear electron acceleration.
\newblock {\em Nature communications}, 2015.

\bibitem{searsmicrobunchiung}
C.~Sears.
\newblock {\em Production, Characterization, and Acceleration of Optical
  Microbunches}.
\newblock PhD thesis, Stanford university, 2008.

\bibitem{Svystun-sim}
E.~Svystun and et. al.
\newblock Beam quality preservation in a laser-plasma accelerator with external
  injection in the context of {EuPRAXIA}.
\newblock {\em Nuclear Instruments and Methods in Physics Research Section A:
  Accelerators, Spectrometers, Detectors and Associated Equipment}, 2017.

\bibitem{Vashchenko-THzgun}
G.~Vashchenko and et. al.
\newblock Performance of the prototype {THz}-driven electron gun for the
  {AXSIS} project.
\newblock {\em Nuclear Instruments and Methods in Physics Research Section A:
  Accelerators, Spectrometers, Detectors and Associated Equipment}, 2017.

\bibitem{Vinatier-possibleexp}
T.~Vinatier and et. al.
\newblock Possible experiments using dielectric-loaded waveguides on the {ARES}
  linac.
\newblock {\em Nuclear Instruments and Methods in Physics Research Section A:
  Accelerators, Spectrometers, Detectors and Associated Equipment}, 2017.

\bibitem{Vinatier-sim}
T.~Vinatier and et. al.
\newblock Simulations of an hybrid and compact attosecond {X}-ray source based
  on {RF} and {THz} technologies.
\newblock {\em Nuclear Instruments and Methods in Physics Research Section A:
  Accelerators, Spectrometers, Detectors and Associated Equipment}, 2017.

\bibitem{andieupraxia}
P.~A. Walker and et. al.
\newblock Horizon 2020 {EuPRAXIA} design study.
\newblock {\em Journal of Physics: Conference Series}, 874(1):012029, 2017.

\bibitem{Zhu-sim}
J.~Zhu.
\newblock {\em Design Study for Generating Sub-femtosecond to Femtosecond
  Electron Bunches for Advanced Accelerator Development at SINBAD}.
\newblock PhD thesis, Hamburg University, 2017.

\bibitem{zhu-compression}
J.~Zhu and et. al.
\newblock Sub-fs bunch generation with sub-10-fs bunch arrival time jitter via
  bunch slicing in a magnetic chicane.
\newblock {\em Phys. Rev. ST Accel. Beams}, 19, 2016.

\end{thebibliography}



\end{document}